\documentclass[12pt,fleqn]{article}
\usepackage{amsmath,amssymb}
\begin {document}
\title{General Non-Applicability of the Liouville Equation in Statistical Mechanics and a New $\mu  - $space Stochastic Equation for Dynamical Systems}
\author{Christopher G. Jesudason}
%\address{Chemistry Department
%University of Malaya,50603 Kuala Lumpur,Malaysia.}
\parindent=0mm
\maketitle
\newtheorem{thm}{Theorem}
\newtheorem{lem}{Lemma}
\newtheorem{cor}{Corollary}

\newtheorem{rem}{Remark}
\newcommand{\R}{{\mathbb R}}

\abstract{\noindent By examining both the divergence of the velocity vector in orthogonal Cartesian coordinate space $\mathbf{\Gamma} $  of dimension   $\R^{\textrm {2fN}}$
 and the structure of the Hamiltonian determining a system trajectory, it is shown that the standard Liouville equation cannot describe anything more than linear motion for typical Hamiltonians with separated momentum and space variables and some significant consequences  such as the Poincare recurrence theorem do not obtain for such Hamiltonians. A new stochastic equation which is everywhere in principle discontinuous is developed for dynamical systems described by a general Hamiltonian which is a functional of the space and momentum variables, and  where the average trajectory of a system point is proven to be orthogonal to any constant energy surface consonant with the system energy at equilibrium. This equation does not assume the presence of binary collision only, as required in the standard first-order Boltzmann equation, and is therefore suitable to describe dense systems as well, and may be viewed as an alternative to the latter. Some new macroscopic variational principles  for non-equilibrium thermodynamical systems are proposed, where one object for future work would be to relate  the microscopic description given here with the macroscopic principles.    }
 
\section {Introduction}
\noindent 
The classical Liouville equation of statistical mechanics is described by trajectories in so-called $\Gamma$ space. In this  $\gamma$ (or $\Gamma$)   space,  N* systems are represented, each with coordinates (\textbf{p},\textbf{q}), where the system has N particles,each of which has f degrees of freedom for the momentum and space variables.This  leads to a 2fN dimensional   $\Gamma$ space where each system may be represented by a point traveling with a velocity $\textbf{v}=(\textbf{\.{p}} ,\textbf{\.{q}})$. From the  definition of the  density-in-space $\rho $ given by $\rho  = \frac{{dN^*}}{{N^*dqdp}}$
 the following consequences obtain: 
\begin{equation} \label{E1}
\frac{{\partial \rho }}{{\partial t}} + { div}(\rho {\bf v}) = 0
\end{equation}
where
\begin{equation} \label{E2}
{ div}\rho {\bf v} = \sum\limits_{j = 1}^{fN} {\left( {\dot q_j \frac{{\partial \rho }}{{\partial q_j }} + \dot p_j \frac{{\partial \rho }}{{\partial p_j }}} \right)}  + \rho \sum\limits_{j = 1}^{fN} {\left( {\frac{{\partial \dot q_j }}{{\partial q_j }} + \frac{{\partial \dot p_j }}{{\partial p_j }}} \right)} 
\end{equation}

%\end{equation}

%\cite{you} 
 The fortuitous cancelation of terms \cite{you} in the 2nd R.H.S. term of Eq.(2) from the relation $\dot q_j  = \frac{{\partial H}}{{\partial p_j }},\dot p_j  =  - \frac{{\partial H}}{{\partial q_j }}$
 involving the Cartesian orthogonal coordinates lead to ${div}\, {\bf v}$ = 0 or 
\begin{equation}\label{E3}
\frac{{\partial \rho }}{{\partial t}} =  - \sum\limits_{j = 1}^{fN} {\left( {\dot q_j \frac{{\partial \rho }}{{\partial q_j }} + \dot p_j \frac{{\partial \rho }}{{\partial p_j }}} \right)}  =  - {\bf v}.grad\rho 
\end{equation}
Eq.~(3) is the Fundamental Liouville equation. 
Because of the fortuitous cancelation of terms involving the momentum and position coordinates, it is \textit{not} immediately apparent that $v_i=constant$ for each component $v_i$ of  $\textbf{v} $ and in fact it  is assumed in general as a  fundamental fact that \textbf{v} is \textit{not} a constant vector, and so this assumption must be tested by testing the divergence independently. Furthermore, the density $\rho (\textbf{p},\textbf{q})$ is sometimes written   independently of the positional coordinates, as in the case of a perfect fluid inside a region with a constant potential, such as in the Gibbs' Canonical ensemble written $\rho ({\bf p}) = C\exp  - \beta H({\bf p},{\bf q})$
 where it cannot be assumed that a  fortuitous cancelation of terms occurs because of the incomplete variables specified,so another method must be utilized to check for the velocity values. Because the Hamiltonian is assumed  {\itshape not} to be explicitly a function of  the time $t$  then the divergence would involve taking limits into regions of $p-q$ space  which are not  constant in energy, whereas  a particular system trajectory would always conserve energy. Hence the macroscopic behavior of  the system predicted from the density is dependent ultimately on the nature of the density flow in $\Gamma$ space, and hence on the divergence of the vector field.   

\section{Test of divergence term and its consequences}
The divergence   is computed from the traditional definition 
\begin{equation}  \label{E4}  
{ div}{\bf J} = \mathop {{\rm lim}}\limits_{\delta V \to 0} \left( {\int\limits_{\partial S} {{\bf J}.{\bf dS}} } \right){\raise0.5ex\hbox{$\scriptstyle {}$}
\kern-0.1em/\kern-0.15em
\lower0.25ex\hbox{$\scriptstyle {}$}}\delta V
\end{equation}
where $\delta V$ is the volume increment  ${\rm \delta }{\bf q}_{\sim k}{\rm \delta }{\bf q}_k   $
 and hence the divergence of the 
 $k^{th}$ component of the coordinate $q_k$ between two parallel faces 1 and 2 at distance ${\rm \delta }{\bf q}_k $
 from each other is  given by 
\begin{equation}\label{E5}
 {div}\left( {\bf J} \right)_k  = 
\left( {\dot q_{{\rm k,2}}  - \dot q_{k,1} } \right){\rm .\delta }{\bf q}_{\sim k}\Bigr/ \delta V 
\end{equation}
where ${\rm \delta }{\bf q}_{\sim k}  $
are the orthogonal surface area increments perpendicular to $\dot q_k $.
 \begin{lem} \label{lm1}
The Liouville expression for the divergence given by Eqs.(\ref{E2}-\ref{E3})\,is equivalent to that derived from computing it from Eq.(\ref{E4}) involving the absolute increments ${\rm \delta }{ q}_k $. 
\end{lem}
\noindent
\textbf{Proof.}
Relative to the mid-point $(p_k,q_k)$, the Hamiltonian at the faces would take  values $
H(p_k ,q_k ) \pm \frac{1}{2}\frac{{\partial H}}{{\partial q_k }}\left| {\delta q_k } \right.
$, with the positive sign for face 2, and the negative for face 1 and therefore Eq.(\ref{E5}) becomes, (where $q_k$ are considered fixed), as 
\begin{eqnarray} 
{ div}\left( {\bf J} \right)_k =  
 \Bigg( \frac{\partial H(p_k ,q_k )}{\partial p_k }\Big|_{{\rm face}\,{\rm 2}}- \frac{\partial H(p_k ,q_k )}{\partial p_k}\Big| _{{\rm face}\,{\rm 1}} \nonumber \\
+  \frac{1}{2}\frac{\partial H(p_k ,q_k )}{\partial p_k \partial q_k } \Big|_{q_{k,2} }  + \frac{1}{2}\frac{\partial H(p_k,q_k )}{\partial p_k \partial q_k } \Big|_{q_{k,1} }   \Bigg) \label{E6}
\end{eqnarray}
Hence  taking limits as  $\delta q_k\rightarrow 0$, leads to
\begin{equation}\label{E7}
div(\dot q)_k   = \frac{{\partial ^2 H}}{{\partial p_k \partial q_k }}
\end{equation}
and an exactly similar derivation for the generalized momenta yields 
\begin{equation} \label{E8}
div(\dot p)_k  =  - \frac{{\partial ^2 H}}{{\partial p_k \partial q_k }}
\end{equation}			
which accords with the standard Liouville expression for the total divergence $div(\textbf{v})=0$, since from Eqs.(\ref{E7}-\ref{E8}), $div({\bf v}) = \sum\limits_{i = 1}^{fN} {(div(\dot q)_k  + div(\dot p} )_k )=0$ but in addition, we can infer 
by  taking limits in the regions without conserving energy in $\Gamma$-space that the individual components are given by Eqs.(\ref{E7}-\ref{E8}) and here, each component may be  identically zero for typical systems, as shown below.We note that  for the Louville equation, \textbf{v} is a non-stochastic, continuous and differentiable vector.
\begin{thm} \label{Th1}
If the Hamiltonian $H$ is partitioned as a linear functional of the coordinates \textbf{q} and the momenta \textbf{p} with no cross-terms as $H = K({\bf p}) + V({\bf q})$
  then each component of the divergence is zero where $div(\dot p)_k  = div(\dot q)_k  = 0$ for all indices $k$.
\end{thm}
\textbf{Proof.}
This follows since $\frac{{\partial ^2 H}}{{\partial p_k \partial q_k }} = \frac{\partial }{{\partial p_k }}\left[ {\frac{{\partial V}}{{\partial q_k }}} \right] = 0$
 because $V$ does not contain any $p_k$ coordinates, and by Maxwellian reciprocity, $\frac{{\partial ^2 H}}{{\partial p_k \partial q_k }} = \frac{{\partial ^2 H}}{{\partial q_k \partial p_k }}.$  For the same reason,$\frac{{\partial ^2 H}}{{\partial p_k \partial q_k }} = \frac{\partial }{{\partial q_k }}\left[ {\frac{{\partial K}}{{\partial p_k }}} \right] = 0$
for the $K$ variable. \linebreak [1]
From Theorem \ref{Th1}  the following obtains:
\begin{cor}
The general solution to the problem of partitioned Hamiltonian coordinates according to Theorem \ref {Th1} is $\frac{{\partial \dot q_k }}{{\partial q_k }} = \frac{{\partial \dot p_k }}{{\partial p_k }} = 0$ for
all k. 
\end{cor}
%\begin{rem}
%\end{rem}
\subsection{Discussion of the solution to Corollary 1} 
The general solution to the problem of Corollary 1 is therefore $\textbf{q}=\textbf{C}t+\beta$ for the position coordinates   and $\textbf{p}=\textbf{D}t+\gamma$  for the momentum coordinates   where   $\textbf{C}$ and $\textbf{D}$ are constant vectors, and $t$ is the time coordinate. The motion is linear with fixed constant forces \textbf{D}, where the velocity vector is written${\bf v} = \left( {\begin{array}{*{20}c}
   {\bf C}  \\
   {\bf D}  \\
\end{array}} \right)$
  . Since $\frac{{\partial \rho }}{{\partial t}} =  - {\bf v}.\nabla \rho $
, it follows that the phase density gradient is orthogonal to the velocity, and such a  solution implies \textit{equidensity}  contours of  $\rho$  that are all parallel to the velocity vector, like a sheath enclosing a line. This result contradicts for instance the well known Gibbs canonical distribution assumption  \cite{Gibbs} which  assumes  that the canonical density $\rho (p,q) = C\exp ( - \beta H)$
 conforms to the Liouville equation together with most other current formulations that imagine $\bf v$ not to be constant. If we consider  a perfect gas with mass m, for instance, then $H = \sum\limits_{k = 1}^{fN} {p_k ^2 /2m} $
 and the stationary solution to the Liouville equation yields$ - \frac{{\beta \rho }}{m}{\bf \dot p}.{\bf p} = 0$. For a canonical ensemble with fixed energy E, the velocity ${\bf \dot p}$
must be tangential to  ${\bf p}$
 , and is therefore tracing the surface of the hypersphere, and is therefore never a constant. It follows that all mathematical theorems that assume the ergodic hypothesis is necessarily false, and in fact the ergodic hypothesis is false from the viewpoint of the Liouville equation with continuous and differentiable variables for the given Hamiltonians. In particular, since the generalized velocities are constant, there is no way in which the Liouville equation can satisfy the celebrated recurrence theorem of Poincar\'{e} and Zermelo with its multifarious and fundamental cosmological consequences, \cite{Terlet}  which states : \textit{The number of phase points, which in their movement leave a given phase volume g without returning into it in the course of time, will be less than any noticeable fraction of the total number of phase points} because  no trajectory may loop back to its initial position because of the  constant velocity solution of systems with Hamiltonians partitioned as in Theorem 1. 

\noindent
 With solutions such as $q_k  = Ct + \beta $
%, $ p_k  = C't + \kappa $
, one might make the identification $\dot q_k \sim\dot p_k /m\sim{\rm constant}$
 , so that only linear motion is described (constant momentum seems implied) which contradicts the normal statistical development, where the Liouville equation is taken as the basis for approximate developments in Fokker-Planck, Boltzmann  and Brownian motion equations \cite{Zwanzig4}. The Boltzmann equation has also been developed from the above Liouville equation \cite{Kirk5} although  Eu \cite{Eu6} fundamentally denies the connection between the Liouville equation and Boltzmann's,  stating  that he used \textit{an irreversible kinetic equation instead  of time-reversal invariant Liouville equation}. It is common knowledge that the internal dynamics of the Boltzmann equation assumes time-reversal invariance, only the macroscopic outcome seems irreversible, and Bolzmann justified this by stating  explicitly  (under the influence of the recurrence theorem of Poincar\'{e} ) that it would take a very long time for the reversal of motion to be observed. It has been shown that in fact time-reversal in general is a false concept from a mathematical point of view\cite{Jesu7}.In fact it is common knowledge that the Boltzmann equation is derivable from the reduced density functions of the Liouville equation \cite{Reich8}. Hence, it is incorrect to suppose that the Liouville equation as it stands is able to provide for the dynamics of a system from which macroscopic properties maybe computed analytically without introducing mathematical discontinuity and stochasticity to the probability density functions, which would lead to the breakdown of the equations themselves. We also note that the quantum version \cite{Reich9} was developed on principles of analogy with the classical result. Finally, Prigogine has also carried out an analysis of the  Liouville equation based  on its analogy with the Schrodinger equation to derive various results; based on the elementary observation above, the analysis is of severely  restricted validity, if not incorrect in general \cite{Prig10}. 
\section{ Development of a  stochastic equation in $\mu$-space}
In view of the above difficulties we develop from first principles another type of equation for describing irreversible phenomena in $\mu$ space, as with the Boltzmann equation.  We recall that $\Gamma$ space represents the set of coordinates for a particular system, implying that the motion of the point represents the evolution of the associated  system. The entire ensemble of systems each with the same Hamiltonian $H_s$ which are coupled to each other to form a Canonical ensemble constitutes one single system called the supersystem with the Hamiltonian $H_S$ which contains all the coordinates of each of the systems j with the same Hamiltonian $H_s$ in $\Gamma '$
superspace; the super-Hamiltonian is therefore a functional of the coordinates 
\begin{equation} \label{E9}
{\bf R} = \left\{ {\left\{ {Q_{ij} } \right\},\left\{ {P_{ij} } \right\}} \right\}
\end{equation}
where the  subscript \textit{i} refers to the particle index $(\textit{i}=1,2,...N)$ for the systems each with $N$ particles, and \textit{j} refers to the system index where $(\textit{j}=1,2,...N')$ for $N'$ systems; $Q$ and $P$ are the spatial and momentum coordinates respectively. In $\Gamma '$
superspace with coordinates $\textbf{R}$ there is only one point which evolves, with the Hamiltonian written as 
\begin{equation} \label{E10}
H_S  = \sum\limits_{j = 1}^{N'} {H_j }  + \sum\limits_{i = 1}^{N'} {} \sum\limits_{j = 1}^{N'} {H_{ij}^{({\rm interact})} } 
\end{equation}
where $H_j$ are the individual system Hamiltonians (with form $H_s$ )containing the coordinates of its particles as variables whereas $H_{ij}^{\textrm{(interact)}}$ contain the (mixed) variables belonging to different systems of the ensemble due to the coupling of the systems; because of the coupling, the system  trajectories in $\Gamma$ space  \textit{can } cross because  the velocity gradients due to the system  Hamiltonian $H_s$ need not equal that due to the supersystem Hamiltonian $H_S$,
i.e.$\left( {\frac{{\partial H_s }}{{\partial {\bf Q}}},\frac{{\partial H_s }}{{\partial {\bf P}}}} \right) \ne \left( {\frac{{\partial H_S }}{{\partial {\bf Q}}},\frac{{\partial H_S }}{{\partial {\bf P}}}} \right)$
for the same set of coordinates $\left\{ {{\bf Q},{\bf P}} \right\}$ belonging to a particular system.The gradients give the incremental trajectory of a system in $\Gamma$ space and clearly relative to an isolated system (with no interactive term) all the system Hamiltonians are the same, implying the same trajectory for any given point in $\Gamma$ space which cannot obtain if interactive terms are present as given in Eq.(\ref{E10}). 
 Single systems of $N$ particles where each particle is interacting with its $N-1$ neighbors may be derived from the above ensemble of $N'$ separate systems by setting $j = 1$.  There is then a reduction from  $\Gamma$ space to the so-called $\mu$ space statistics of the single system, where the Non-Louville density ${\rm D}^{{\rm N - L}} $
would give the normalized number of particles within the volume element $\mathbf{\Delta p\Delta q}$ where $(\mathbf{p,q})$ are the momentum and position coordinates common to all the particles, but in  general, when $j\not=1$, there is a reduction from $\Gamma'$ superspace to $\Gamma$  system space. Since trajectories can cross at any point in  $\Gamma$ space, the simplified Liouville equation of Eq.(\ref{E2}) or other methods which employ continuous functions  with non-probabilistic derivatives cannot be employed directly.  For an ensemble of systems interacting with one another where the coordinates have been partitioned (with respect to the super-Hamiltonian and the coordinates that belong to a particular system, as in Eqn.(9) so that each system is a point in  $\Gamma$  space, then the non-Liouville density ${\rm D}^{\textrm{N-L}}$ may be expressed  generally as 
\begin{eqnarray}
 \frac{{{\rm dD}^{{\rm N - L}} {\rm (P}_{\rm 1} {\rm .,}..{\rm P}_{\rm N} {\rm .,Q}_{\rm 1} {\rm .,}..{\rm Q}_{\rm N} {\rm .,t)}}}{{{\rm dt}}} 
& = & \frac{{\partial {\rm D}^{{\rm N - L}} }}{{\partial {\rm t}}}{\rm   + }\frac{{\partial {\rm D}^{{\rm N - L}} }}{{\partial {\bf Q}}} \cdot {\bf \dot Q}  
  + {\rm  }\frac{{\partial {\rm D}^{{\rm N - L}} }}{{\partial {\bf P}}} \cdot {\bf \dot P} \nonumber\\ 
& = & {\rm K(P}_{\rm 1} {\rm .,}..{\rm P}_{\rm N} {\rm .,Q}_{\rm 1} {\rm .,}..{\rm Q}_{\rm N} {\rm .,t)} \label{E11} 
 \end{eqnarray}

The velocities $({\bf \dot P},{\bf \dot Q})$
 cannot be derived from the Liouville equation and so one must be able to independently determine the meaning of K as well as $({\bf \dot P},{\bf \dot Q})$
: Clearly, Eq.(\ref{E11}) would represent in  general form an evolution equation. Clearly, there is no immediate relation or identity between the density $\rho=\textrm{D}$ of the original Liouville Eq. (\ref{E3}) and the above ${\rm D}^{{\rm N-L}}$. 
 The P's and Q's of this interacting system are the independent variables of ${\rm D}^{{\rm N-L}}$, and one cannot use $H_s$  (pertaining to a particular particle $s$ since each particle is its own system)  to derive velocities directly
, as is sometimes assumed in the Liouville description. The non-Liouville density ${\rm D}^{{\rm N-L}}$ is given for N* interacting systems, each with N particles having coordinates $\rm{(P_a', Q_a')}$ as 
\begin{eqnarray}
{\rm D}^{{\rm N - L}} {\rm  (Q_a',P_a')  }& = &{\rm    }\sum\limits_{\rm i} {\sum\limits_{\rm j} {\frac{{\rm  (f(P_{ij}) - P_a')  }{{\rm (f'(Q_{ij}) - Q_a')}}}{{{\rm NN^*}}}{\rm dP}_{\rm a}^{\rm '} {\rm dQ}_{\rm a}^{\rm '} } }\nonumber \\
{\rm where} \,\,\,\,{\rm  (f(P_{ij}) - P_a')}& = & 1\,\,\,{\rm if}\,\,{\rm f(P}_{{\rm ij}} ) - {\rm P}_{\rm a}^{\rm '}  \in {\rm dP}_{\rm a}^{\rm '} \,{\rm and}\,\,{\rm zero}\,\,{\rm otherwise}\,\,{\rm and}\nonumber\\
{\rm  (f'(Q_{ij}) - Q_a')}& = & 1\,\,\,{\rm if}\,\,{\rm f'(Q}_{{\rm ij}} ) - {\rm Q}_{\rm a}^{\rm '}  \in {\rm dQ}_{\rm a}^{\rm '}\,{\rm and}\,{\rm zero}\,{\rm otherwise} .\label{E12}
\end{eqnarray}

 For the above Eq.(\ref{E12}) the  primed coordinates belong to   $\Gamma$ space and the unprimed to  ${\rm\Gamma'}$  superspace. The mapping function $
{\rm f(P}_{{\rm ij}} {\rm )}$
 maps the coordinate  ${\rm P}_{{\rm ij}} $
 of ${\rm\Gamma'}$   space   to the  $\Gamma$ space $
{\rm P'}_{\rm i} $, and ${\rm dP'} $
   is the volume element of arbitrary size (set by a particular algorithm which depends on the physical dimensions of the particles under study of the momentum space, with similar definitions  for  the ${\rm f'}$ mapping function which acts on the  ${\rm Q'}$ coordinates). We note that no Dirac delta functions are immediately implied by Eqn.(\ref{E12}).  The mapping induced by Eq.(\ref{E12}) does not in any obvious way allow the Q and P variables of $\Gamma'$  space  to be continuously transformed to the ${\rm Q'}$ and ${\rm P'}$  variables of $\Gamma$ space (which are variables of ${\rm D^{N-L}}$).  Such precise mapping functions must be outlined before progress may be made with an extended form of the Liouville equation which may detail  full irreversible behavior. We now present a possible approach for the simplest case  of  an equation describing irreversible behavior in $\mu$ space leading to an alternative to the   Boltzmann equation which is  founded on elementary reasoning of stochastic processes, which does not accord completely with :
\begin{itemize}
\item[(a)]
authors who   disavow the use of  equations of the Liouville kind \cite{Eu11}, because we are able to derive a form which resembles the Liouville equation, but the variables can only be defined probabilistically
\item[(b)]
the utilization of the collision integral \cite{Eu12}    because of  the problems already mentioned concerning the use of time-reversibility in the collision integral which are mathematically incorrect
\end{itemize}
We resort to a statistical method which can be in principle  verified computationally for all quantities mentioned, but where stochastic differentials only are involved,  and where  a stochastic calculus can easily be elaborated as needed to any degree of accuracy. 

\begin {thm} \label{th2}
It is possible to derive a stochastic equation having the form of the Liouville equation which is more general than the standard  Boltzmann equation in $\mu$-space without recourse to any of the reversibility assumptions implicit in the Boltzmann equation method.
\end{thm}
\textbf{Proof.}
We  use the superensemble method described in Eqs.(\ref{E11}-\ref{E12}), with the special case of  Eq.(\ref{E9}) with j=1, where the super-Hamiltonian 
$H_{\Gamma ^{'} }$ 
in $\Gamma'$ space (represented by one point in this 6N dimensional space) would represent the motion of  the N particles having the generalized Hamiltonian $H_{\Gamma ^{'} }{(P_1,P_2,....P_{3N}, Q_1,Q_2,....Q_{3N})} $
 with the {\bf P} momentum and {\bf Q} spatial coordinates with subscripted labels referring to the particles; the potential is completely arbitrary in the coordinates given, except that the Hamiltonian is assumed conservative; an example of this system is  a collection of $N$ particles for ${\rm\Gamma'}$ space, whereas the $\mu$ space consists of the $(\bf p, \bf q)$ coordinates common to all the particles.  The $\mu$ space Cartesian coordinates will be $p_x, p_y$,and $p_z$,  for the momentum  and $q_x, q_y$ and $q_z$  for the spatial components. Then for  finite increments $\Delta q_i, \Delta p_i(i = x,y,z)$,which are components of the respective vectors ${\bf \Delta q}$ and ${\bf \Delta p}$,  the unnormalized $\mu$ density-in-phase $F(\bf p,\bf q)$ representing the number of particles within the volume element $\Delta\bf p\Delta\bf q$ would be such that the following equations obtain:
 $$\Delta p_x \Delta p_y \Delta pz\Delta q_x \Delta q_y \Delta q_z \rm F({\bf p},{\bf q})
= \rm M(\Delta {\bf p},\Delta {\bf q},({\bf P},{\bf Q})_{\Gamma ^{'}},({\bf p},{\bf q})_\mu)\nonumber $$
$$= \sum\limits_{j = 1}^N {\delta '(p_x  - P_{x,j} } ).\delta '(p_y  - P_{y,j} ).\delta '(p_z  - P_{z,j} ).\delta '(q_x  - Q_{x,j} ).\delta '(q_y  - Q_{y,j} ).\delta '(q_z  - Q_{z,j} ) $$
 $$= \sum\limits_{j = 1}^N {\prod\limits_{i = x,y,z} {\delta '(p_i  - } } P_{i,j} )\delta '(q_i  - Q_{i,j} )$$
\begin{eqnarray} \nonumber
{\rm where}\,\delta '(p_i  - P_{i,j} ) &=& 1\,\,{\rm if}\,\left| {p_i  - P_{i,j} } \right| < \Delta p_i \,{\rm and}\,{\rm zero}\,{\rm otherwise,and}\nonumber\\
\delta '(q_i  - Q_{i,j} ) &=& 1\,\,{\rm if}\,\left| {q_i  - Q_{i,j} } \right| < \Delta q_i \,{\rm and}\,{\rm zero}\,{\rm otherwise.}\label{E13}
\end{eqnarray}
The  set of incremental limits
$\left\{ {\Delta {\rm p}_{\rm i} ,\Delta q_j } \right\}$
 are dependent on the particle characteristics and   must be chosen as fixed quantities in any numerical simulation. Since $\rm M$ in an integer function, it is very clear in general that  the limit $\mathop {\lim }\limits_{\delta t \to 0} \frac{{\delta F}}{{\delta t}}$
does not  exist (e.g. as in the Wiener process), so that  only a ratio based on finite values of the independent variables can  be defined in the stochastic calculus for finite time increments  $0<\delta t <T$ for some specified T which is system dependent. To develop the above density , we need to define various quantities averaged over time T.
 
\noindent 
{\itshape Determination of average quantities}:The subscripted variables after the vertical bar refer to the variables that are kept constant in the averaging expression.\linebreak[1]
$\left\langle \frac{{\delta F}}{{\delta t}}\left\vert_{{\bf p},{\bf q}}\right.
    \right\rangle$:instead of directly differentiating (\ref{E13}) we write instead\linebreak[1]
\begin{eqnarray}
 {\bf \Delta p\Delta q}\left\langle {\frac{{ \delta F}}{{\delta t}}\left|_ {{\bf p,q}} \right.} \right\rangle  &=& \frac{{\left\{ { - {\rm M}({\bf p},{\bf q},{\bf P}{\rm ,}{\bf Q}{\rm ,t)} + {\rm M}({\bf p},{\bf q},{\bf P} + {\bf \dot P}{\delta t,}{\bf Q} + {\bf \dot Q}{\rm \delta t,t} + {\rm \delta t)}} \right\}}}{{{\rm \delta t}}} \nonumber\\ 
  &=& \left\langle {\frac{{ \delta M}}{{ \delta t}}\left|_ {{\bf p},{\bf q}} \right.} \right\rangle \label{E14}
\end{eqnarray}
where\,\,$ {\bf \dot P} =  - \frac{{\partial {\rm H}_{{\rm \Gamma '}} }}{{\partial {\bf Q}}}{\rm ,}{\bf \dot Q} = \frac{{\partial {\rm H}_{{\rm \Gamma '}} }}{{\partial {\bf P}}}$. Eq.(\ref{E14})  is can be simplified using stochastic averaging, which will not be pursued here.  
Similarly, we may define for the momentum coordinates
\begin{eqnarray}
 {\bf \Delta p\Delta q}\left\langle {\frac{{ \delta F}}{{\delta p_i}}\left|_ {t,{\bf q}} \right.} \right\rangle  &=& \frac{{\left\{ { - {\rm M}({ p_i},{\bf q},{\bf P}{\rm ,}{\bf Q}{\rm ,t)} + {\rm M}(p_i+\delta p_i,{\bf q},{\bf P},{\bf Q},t  {)}} \right\}}}{{ \delta p_i}} \nonumber\\ 
  &=& \left\langle {\frac{{ \delta M}}{{ \delta p_i}}\left|_ {t,{\bf q}} \right.} \right\rangle \label{E15}
\end{eqnarray}							
and a similar expression for the spatial coordinates where 
%\begin{eqnarray}
 $${\bf \Delta p\Delta q}\left\langle {\frac{{ \delta F}}{{\delta q_i}}\left|_ {t,{\bf p}} \right.} \right\rangle $$

 $$= \frac{{\left\{ { - {\rm M}({\bf p},{ q_i},{\bf P}{\rm ,}{\bf Q}{\rm ,t)} + {\rm M}({\bf p},q_i+\delta q_i,{\bf P},{\bf Q} {\rm ,t}  {\rm )}} \right\}}}{{ \delta q_i}}$$
\begin{equation}
  = \left\langle {\frac{{ \delta M}}{{ \delta q_i}}\left|_ {t,{\bf p}} \right.} \right\rangle \label{E16}
\end{equation}
However, Eqs.(\ref{E15}-\ref{E16}) will not be further considered here. We define the vector component $v_i$ of the average velocity {\bf v} of the particles  contained within the $\mu$ space volume element ${\bf \Delta p\Delta q}$ as $v_j  = \frac{{\sum\limits_{l = 1}^{N'} {\dot p_{j,l} } }}{{N'}}$  for momentum and $v_i  = \frac{{\sum\limits_{l = 1}^{N'} {\dot q_{i,l} } }}{{N'}}$for space variables where the subscripts $i$ or $j$ take coordinate labels  $(x,y,z)$ , and $l$ refers to the particle label where $P_{i,l}\equiv p_{i,l}$ and  $ Q_{i,l}\equiv q_{i,l}$ whenever $
( p_{i,l} , q_{i,l} ) \in {\bf \Delta p\Delta q}$, and where $N' = \rm F{\bf \Delta p\Delta q}$.
Since particle numbers are conserved we can write
\begin{eqnarray} 
 \left\langle \frac{{\delta F}}{{\delta t}}\left\vert_{{\bf p},{\bf q}}\right.
    \right\rangle&=&- \nabla .\rho \sum\limits_i {{\rm\bf v}_i } \nonumber\\
&=&-\overline {\nabla .\rho \sum\limits_i {{\rm\bf v'}_i }}\label{E17}
\end{eqnarray}
where in the right hand side of  Eq.(\ref{E17}), the bar denotes the average of all the particle flow vectors $\rho {\bf v'}_i$, where $i$  represents a particle within the $\mu$-space volume increment  $\Delta {\bf p}\Delta {\bf q}$ and the non-Liouville density $D^{N-L}$is given by $D^{N-L}=\rho\bigl/N\equiv \rm F(\bf p,\bf q,t)\bigl/$$N$ where $N$ is the total number of particles in the system. Eq.({E17}) then gives 
\begin{equation}
\left\langle \frac{\delta F}{{\delta t}}\left\vert_{{\bf p},{\bf q}}\right.\right\rangle
  + \nabla \rho  \cdot \sum\limits_i {{\bf v}_i }  + \rho \nabla  \cdot \sum\limits_i {{\bf v}_i }  = 0. \label{E18}
\end{equation}
Each particle $i$ within the volume element $\Delta {\bf p}\Delta {\bf q}$ must also have   some coordinate $(P_j,Q_j)$ in $\Gamma'$ space, so where the exact $i$ particle trajectory is concerned, we can write the following entities:\linebreak[1]
$\frac{\partial }{{\partial p_i }} \equiv \frac{\partial }{{\partial P_j }}$ and
$\frac{\partial }{{\partial q_i }} \equiv \frac{\partial }{{\partial Q_j }}$ 
and hence for any pair of velocities , the divergence terms  cancel by the Hamiltonian result in Eqn.(\ref{E3}), and so $\nabla.{\rm\bf v}_i =0$ and thus $\nabla.\sum_{i}{\rm\bf v}_i =0$ and  $\nabla\rho.\sum_{i}{\rm\bf v}_i= \nabla\rho.{\rm\bf v}$ where $\bf v$ is the average velocity of the non-liouville density function $\rho\equiv F({\rm\bf p},{\rm\bf q},t)$. {\itshape Thus, we have the remarkable stochastic equivalent of the Liouville equation}:
\begin{equation}
\left\langle {\frac{{\delta { F}}}{{\delta {\rm t}}}\left| {_{{\bf p,q}} } \right.} \right\rangle  =  - \nabla \rho .{\bf v} \label{E19}
\end {equation}
and at equilibrium, we infer $
\left\langle {\left. {\frac{{\delta F}}{{\delta t}}} \right|_{{\bf p},{\bf q}} } \right\rangle  = 0 \Rightarrow  - \nabla \rho .{\bf v} = 0$.
\begin{rem}
It is expected therefore that  the average velocity trajectory of the particles would be orthogonal to the gradient of the density at equilibrium. This analysis does not support the Poincar\'{e} theorem because there will be an infinite number of  trajectories with different gradients for each point in $\mu$-space and therefore  for  any arbitrary phase volume $\Delta {\bf dp},{\bf dq}$,   no limit to a unique trajectory exists.
\end{rem}
 On account of the importance of the above equation of motion, we shall refer to it as the evolution equation for the {\itshape  $\mu$-space microcanonical system  trajectory}. The validity of this  equation does not depend on any regime of particle density, nor on any dubious assumptions regarding Newtonian dynamics such as demanded by the Boltzmann {\itshape \textquotedblleft Stosszahlansatz\textquotedblright}; it  therefore can be viewed as an  
alternative to the Boltzmann equation, and may be directly tested by computer simulation, and it would be of urgent interest to do this  for simple cases, for instance hard spheres contained in a fixed volume at constant energy. The obvious generalization of the above would  be to write down a stochastic equation for the $\mu$-space evolution of a system connected to others in a canonical ensemble; this evolution equation would constitute the {\itshape $\mu$-space canonical system  trajectory} and lastly, and not so obviously, it may be possible to write down an evolution equation for a system exchanging both energy and matter (zero-particle conservation) leading to the evolution equations for the {\itshape $\mu$-space Grand-Canonical system  trajectory }, and these   evolution equations  can be tested by computer simulations. Eq.(\ref{E13}) provides a means to test the Boltzmann H-Function according to this formulation for the non-Liouville particle distribution function. \linebreak[4]

\noindent
Lastly, for such complex systems, it would be of interest to
link the general equations above to variational principles that have been proposed and applied to irreversible thermoelectrical and other \cite {Jesu13,Jesu14}   non-equilibrium systems which pertain to both (a){\itshape the dynamical flow vectors} and (b) {\itshape the stationary field and matter distribution of the system, by analogy with Gibbs' stationary principles for thermostatic equilibrium}.
It will be one  purpose of future research to correlate the present work to the non-linear  variational principles for non-equilibrium processes mentioned above where (a) has the form
\begin {equation} \label{E20}
\int\limits_{\partial C} {dS = 0} 
\end{equation}
and (b) the form
\begin {equation} \label{E21}
\delta \left[ {\int_{\partial C} {dS} } \right]_{U,\left\{ {\bf x} \right\}}  = 0.
\end{equation}
In Eq.(\ref{E20}), $dS$ represents the entropy change for a \textquotedblleft disintegrating\textquotedblright  or \textquotedblleft recoverable\textquotedblright system \cite{Jesu13} along an actual trajectory, whereas Eq.(\ref{E21}) refers to the entropy change $dS$ of a collection of particles - {\itshape the macroparticle}-  subjected to a virtual perturbation in a thermofield subjected to constraints  $\left\{ {\bf x} \right\}$ pertinent for the entire system (such as particle number and volume of the entire system, but not necesarily the macroparticle). Another topic for further research is the  verification of the above proposed principles by direct computation over model systems.\linebreak [4]
\textbf{Acknowledgements}:I wish to thank (a) V. Lakshmikantham (F.I.T,{\itshape U.S.A.})\,for organising WCNA-2000,\,(b)A. Sengupta (I.I.T (Kanpur), {\itshape India}) for the invitation to the conference and (c) University of Malaya, {\itshape  Malaysia} for the financial assistance.

\noindent
Christopher G. Jesudason \\
Lecturer \\
Chemical Physics Division,Chemistry Department \\
University of Malaya \\
50603 Kuala Lumpur, Malaysia \\
email: jesu@kimia.um.edu.my  \\[1mm]
      
\end {document}